\documentclass[aip,reprint]{revtex4-2}

\draft 
\usepackage{xspace}
\usepackage{esvect}
\usepackage{xcolor}
\usepackage{graphicx}
\usepackage[table,xcdraw]{colortbl}
\newcommand{\STO}{SrTiO\textsubscript{3}\xspace} 
\newcommand{\LAO}{LaAlO\textsubscript{3}\xspace} 
\newcommand{\GAO}{$\mathrm{\gamma}$-Al\textsubscript{2}O\textsubscript{3}\xspace} 

\newcommand{\RRT}{$R^{RT}_s$\xspace}

\newcommand{\STOn}{SrTiO\textsubscript{3}}

\renewcommand{\u}[1]{\textsuperscript{\textit{#1}}}
\newcommand{\ur}[1]{\textsuperscript{#1}}
\renewcommand{\d}[1]{\textsubscript{\textit{#1}}}
\newcommand{\dr}[1]{\textsubscript{#1}}
\newcommand{\mob}{$\mathrm{cm^2/Vs}$\xspace}

\newcommand{\gmu}{$\mathrm{\mu}$\xspace} 

\newcommand{\mgmu}{$\mathrm{\mu}$} 

\definecolor{Brown}{HTML}{efdfd7}

\renewcommand{\textcolor}[1]{}
\begin{document}


\title{Leveraging high fluence and low pressure for pulsed laser deposition of high-mobility \GAO/\STO heterostructure growth} 


\author{Thor Hvid-Olsen}
 \altaffiliation{These authors contributed equally.}
 \affiliation{Department of Energy Conversion and Storage, Technical University of Denmark, Fysikvej 310 DK-2800 Kgs. Lyngby, Denmark.}
\author{Christina Hoegfeldt}%
 \altaffiliation{These authors contributed equally.}
 \affiliation{Department of Energy Conversion and Storage, Technical University of Denmark, Fysikvej 310 DK-2800 Kgs. Lyngby, Denmark.}%
\author{Amit Chanda}
 \affiliation{Department of Energy Conversion and Storage, Technical University of Denmark, Fysikvej 310 DK-2800 Kgs. Lyngby, Denmark.}%

\author{Alessandro Palliotto}%
 \affiliation{Department of Energy Conversion and Storage, Technical University of Denmark, Fysikvej 310 DK-2800 Kgs. Lyngby, Denmark.}%
\author{Anshu Gupta}%
 \affiliation{Department of Energy Conversion and Storage, Technical University of Denmark, Fysikvej 310 DK-2800 Kgs. Lyngby, Denmark.}%
\author{Dae-Sung Park}%
 \affiliation{Department of Energy Conversion and Storage, Technical University of Denmark, Fysikvej 310 DK-2800 Kgs. Lyngby, Denmark.}%
\author{Thomas Sand Jespersen}%
 \affiliation{Department of Energy Conversion and Storage, Technical University of Denmark, Fysikvej 310 DK-2800 Kgs. Lyngby, Denmark.}%

\author{Felix Trier}
 \altaffiliation{Correspondng author.}
 \email{fetri@dtu.dk}
 \affiliation{Department of Energy Conversion and Storage, Technical University of Denmark, Fysikvej 310 DK-2800 Kgs. Lyngby, Denmark.}%


\date{\today}

\begin{abstract}
High-mobility oxide heterostructures could be applied for high-frequency devices, transparent conductors, and spin-orbit logic devices. \STO is one of the most studied oxide substrate materials for heterostructures. To date, the highest \STO-based charge carrier mobility at 2 K was measured in the interfacial 2-dimensional electron gas (2DEG) of \GAO/\STO. The formation mechanism and origin of the high electron mobility are not yet fully understood. This investigation presents a successful growth protocol to synthesise high-mobility \GAO/\STO interfaces, and a description of the underlying growth optimisation. Furthermore, indicative features of high-mobility \GAO/\STO, including the room-temperature sheet resistance, are presented. Signs of epitaxial and crystalline growth are found in a high-mobility sample ($\mu^\mathrm{10K} = 1.6 \times 10^4$ cm\ur{2}/Vs). Outlining the growth mechanisms and comparing 40 samples, indicates that low pressure ($P\approx 1 \times 10^{-6}$ mbar) is essential for high-mobility \GAO/\STO interfaces. \GAO having single-element cations allows higher laser fluences during growth, compared to thin films with multi-element cations such as \LAO, without causing stoichiometric imbalances.
\end{abstract}


\maketitle 

\section{Introduction}
For two decades, oxide-based two-dimensional electron gasses (2DEGs) have been intensively studied\cite{Ohtomo2004} for their diverse features, such as superconductivity,\cite{Reyren2007} ferromagnetism,\cite{Brinkman2007} glassy dynamics,\cite{Maity2025}, tunable Rashba spin-orbit coupling,\cite{Caviglia2010}  high-mobility transport,\cite{Chen2013,Christensen2018} extreme magnetoresistance,\cite{Christensen2024} and spin-charge interconversion.\cite{Trier2020, Noel2020, Vaz2020} In recent years, several steps have been taken toward implementing such oxide interfaces. This includes the generation and detection of spin-currents in ferromagnet-free oxide nanodevices\cite{Trier2020} and fabrication of free-standing oxide interface membranes,\cite{Sambri2020, Dahm2021, Li2022}, paving the way for cointegrating oxide and silicon electronics and spintronics in everyday devices. While most studies focus on the \LAO/\STO interface, the highest charge carrier mobility, $\mu = 140\,000$ \mob at 2 K, is recorded for \GAO/\STO,\cite{Chen2013} presumably due to a band-inversion that renders the \textit{d}\d{xz/yz} band dominant in electronic transport.\cite{Cao2016,Chikina2021}  This attractive property allows for mean free path dependent quantum and spintronics measurements,\cite{Oltscher2014} with potential in applications including high-frequency devices,\cite{Trier2018} transparent conductors,\cite{Trier2018} and spin–orbit logic devices.\cite{Manipatruni2019}\\
At room temperature, longitudinal optical (LO) phonons and electron-electron scattering dominate the carrier mobility in \STO.\cite{Christensen2018, Himmetoglu2014, Mikheev2015} These two scattering phenomena dominate at different carrier density ranges. 
One attempt to counteract the LO phonon scattering, leveraged the non-polar, low effective electron mass, oxide BaSnO\dr{3}, instead of \STO, to reach room temperature mobilities on 60 \mob, an order of magnitude higher than \STO-based 2DEGs.\cite{Eom2022} The different mobilities of the two materials are attributed to the orbitals near the Fermi Energy. In \STO, these orbitals are the \textit{d}-band orbitals of Ti, with degenerate \textit{t}\dr{2g} symmetry, while it is the isotropic \textit{s}-band orbitals in  BaSnO\dr{3}.  From the less dispersive \textit{d}-band orbitals, the carriers have a higher rate of inter-orbital scattering in \STO compared to \textit{s}-band orbital BaSnO\dr{3}.\cite{Eom2022}
\begin{figure*}
    \begin{center}
    \vspace{0pt}
    \includegraphics[width=1\textwidth]{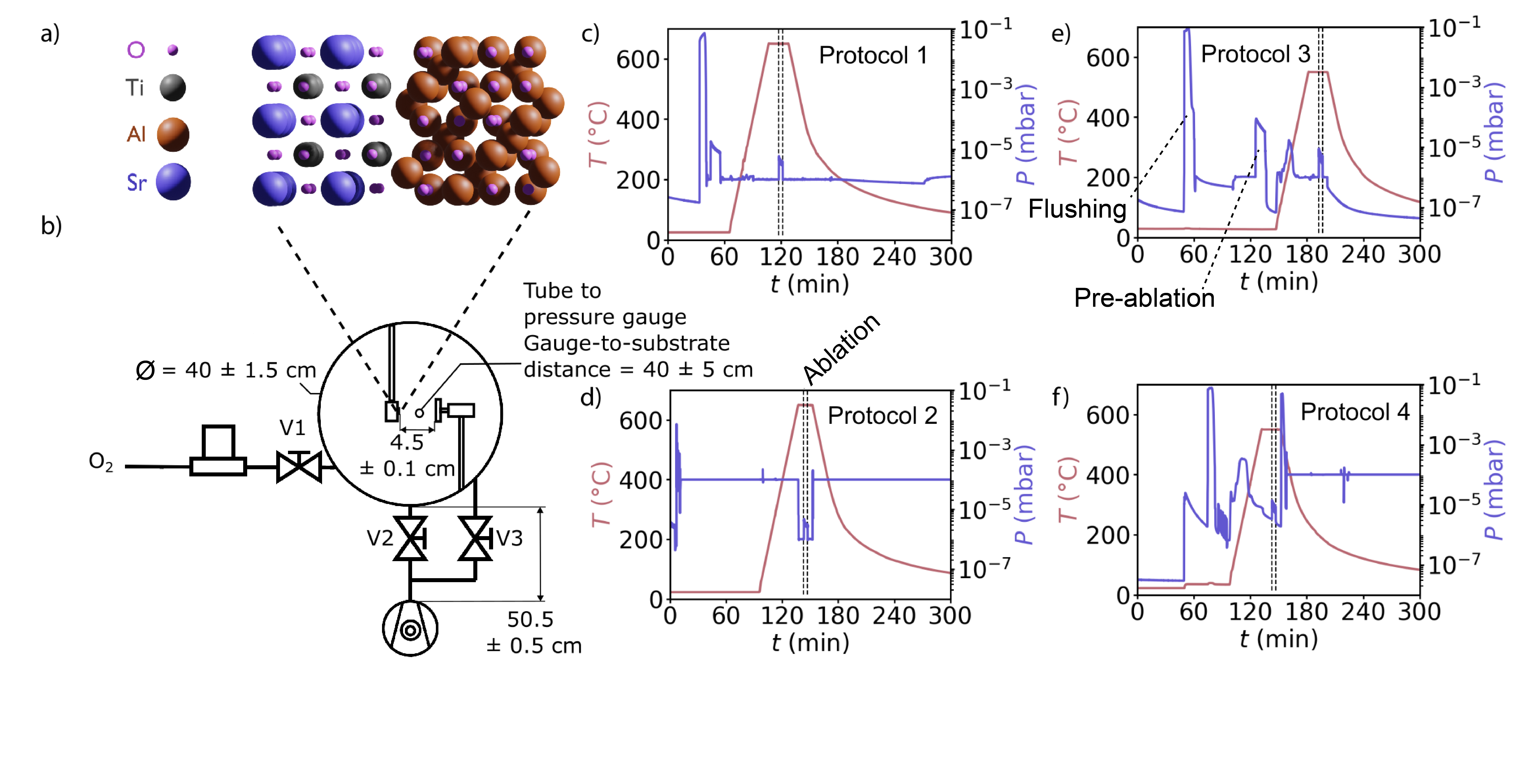}
    \caption{\textbf{Pressure-control for increased reproducibility and crystal structure.} a) Schematic of the \GAO/\STO crystal structure grown with the PLD setup illustrated in b). c-f) Examples of temperature (brown) and pressure (purple) profiles on Protocols 1-4, respectively.
    \label{Fig1}}
    \end{center}
\end{figure*}
\noindent
At low temperatures, phonon-scattering vanishes, and the main mobility-reducing mechanism in \STO 2DEGs is impurity scattering. This complicates achieving high mobility, as one of the major mechanisms forming the 2DEGs uses oxygen vacancies. Oxygen vacancies act as both dopants and scattering sites, motivating several studies into the spatial separation of oxygen vacancies and the 2DEG that arises from them.\cite{Zurhelle2020, Chikina2021, Chen2015, Huijben2013} \\
Here, we present an optimisation of the pulsed laser deposition (PLD) parameters of \GAO grown on \STO towards forming a high-mobility 2DEG. First, the deposition chamber used in growth, along with four different protocols is presented. Subsequently, relations between low-temperature mobility and residual resistance ratio (\textit{RRR}) and room-temperature sheet resistance ($R_S^{RT}$), respectively, are analysed. Then the interface structure of a high-mobility sample is addressed, including epitaxial strain.  We then discuss the mechanisms responsible for the high-mobility interfaces throughout the growth process, from as-received substrate \STO to high-mobility \GAO/\STO(001) 2DEG. Based on this discussion, we compare the growth protocols and highlight indications of a strong pressure-dependence of the 2DEG formation.\\
\section{Experiment}
Commercial \STOn(100) substrates\cite{Shinkosha2024} were cleaned by ultrasonication in acetone (480 s), isopropanol (IPA,480 s), and acetone:IPA 50:50 (480 s), then annealed in a 1 bar O\dr{2} atmosphere with a 60 min dwell time at 1000$^\circ$C and an up/down ramp rate of 100 $^\circ$C/h ($\approx$ 25 h in total with passive cooling). Before \GAO deposition, the substrates underwent a second round of ultrasonication in acetone (480 s), IPA (480 s), and acetone:IPA 50:50 (480 s), respectively. The depositions, carried out by PLD, used a single-crystalline $\alpha$-Al\dr{2}O\dr{3} target. The ablated plasma species forms the \GAO structure on \STO, as displayed in Fig. \ref{Fig1}a). A 248 nm KrF excimer laser with a repetition rate of 1 Hz was used. We find that the detailed configuration of the growth reactor plays an important role in the optimisation process. The PLD chamber, illustrated in Fig. \ref{Fig1}b), is directly connected to three manually controlled valves. One of the valves, V1, is connected to a mass flow controller for oxygen injection, while two other valves, V2 and V3, connect the chamber to a turbomolecular pump. This pump is further connected to a backing pump through another manual valve (not shown). V2 and V3 controlled the specific pressure when working below 10\ur{-4} mbar. V2 is a manual gate valve controlling the range of pressure, while V3 is an angle valve, allowing for fine-tuning of the pressure. V1 was open during O\dr{2} flushing and automatised proportional–integral–derivative (PID) pressure control. The thermocouple, measuring the sample temperature, resides in the heater centre, and the substrate-target distance was 4.5 $\pm$ 0.1 cm.\\
\begin{table*}
\caption{\label{T:params}\textbf{Overview of growth parameters.}}
\begin{tabular}{lccccc}
\hline
\hline
Parameters & Abbreviaion & Protocol 1 & Protocol 2 & Protocol 3 & Protocol 4\\
\hline
Flushing Pressure (mbar) & $P_{\mathrm{flush}}$ & 7 $\times$ 10\ur{-2} & NA & 7 $\times$ 10\ur{-2} & 7 $\times$ 10\ur{-2}\\
Deposition pressure (mbar) & $P_{\mathrm{dep}}$ & 1-2 $\times$ 10\ur{-6} & 1 $\times$ 10\ur{-6} & 1 $\times$ 10\ur{-6} & 1 $\times$ 10\ur{-6}\\
Preablation Pressure (mbar) & $P_{\mathrm{pre}}$ & $P_{\mathrm{dep}}$ & $P_{\mathrm{dep}}$ & $P_{\mathrm{dep}}$ & $P_{\mathrm{dep}}$\\
Heating pressure (mbar) & $P_{\mathrm{heat}}$ & 1 $\times$ 10\ur{-6} & 1 $\times$ 10\ur{-4} & Base $P$ & Base $P$\\
Cooling pressure (mbar) & $P_{\mathrm{cool}}$ & 1 $\times$ 10\ur{-6} & 1 $\times$ 10\ur{-4} & Base $P$ & 1 $\times$ 10\ur{-4}\\
Ramp rate ($^\circ$C/min) & $\Gamma_{\mathrm{ramp}}$ & 15 & 15 & 10,15 & 10,15\\
Deposition temperature ($^\circ$C) & $T_{\mathrm{dep}}$ & 650 & 550, 650 & 550 & 550, 660, 650\\
Preablation time (min) & $t_{\mathrm{pre}}^{\mathrm{dep}}$ & 10 & 10 & 10 & 10\\
Predeposition annealing time (min) & $t_{\mathrm{pre}}^{\mathrm{ann}}$ & 11,16,51 & 6-51 & 11 & 11\\
Postdeposition annealing time (min) & $t_{\mathrm{post}}$ & 5 & 5 & 5 & 5\\
Laser fluence (J/cm\ur{2}) & $F$ & 3.1-9.6 & 3.7-9.2 & 7.1-9.3 & 6.4-8.0\\
\hline
\hline
\end{tabular}
\end{table*}
Pressure and temperature data of four protocols are displayed in Fig. \ref{Fig1}c-f) and the different growth parameters are listed in Table \ref{T:params}. Protocol 1 was the first attempted protocol that yielded high-mobility samples, the other three protocols were attempts to automate and further optimise the growth process concerning reproducibility and mobility. \\
Protocol 1, Fig. \ref{Fig1}c), started with flushing the chamber with $P_{\mathrm{flush}}$ = 7 $\times$ 10\ur{-2} mbar O\dr{2} to ensure that the dominant gas in the chamber was O\dr{2}. Then the chamber was evacuated to $P_{\mathrm{pre}}$ = 1-2 $\times$ 10\ur{-6} mbar and the target preablated for $t_{\mathrm{pre}}^{\mathrm{dep}}$ = 10 min with a shutter shielding the \STO substrate. Using $P_{\mathrm{heat}}= P_{\mathrm{cool}}=P_{\mathrm{pre}}$, the heater, onto which the substrate was thermally and electrically attached using conducting silver paint, was heated to the deposition temperature of $T_{\mathrm{dep}}$ = 650 $^\circ$C with a ramp rate of $\Gamma_{\mathrm{ramp}}$ = 15 $^\circ$C/min. The sample was pre-annealed for $t_{\mathrm{pre}}$ = 11 min, except where else is stated, at deposition pressure and temperature before 4 min of ablation began. After ablation, the sample was post-annealed for $t_{\mathrm{post}}$ = 5 min at deposition pressure and temperature before the onset of the cooling. The cooling was, like the heating, set to a ramp rate of $\Gamma_{\mathrm{ramp}}$ = 15 $^\circ$C/min, below 300 $^\circ$C the ramp rate decreased since the sole cooling power came from the local, low-pressure, environment.\\
All samples grown with protocol 1 were attached to the heater through a metal plate. For the subsequent protocols, the omicron plate was removed to obtain a more precise representation of the sample temperature. Supplementary Material Sec. IV describes the growth of a sample using Protocol 1 without the metallic plate. This resulted in a bulk conducting sample, due to the additional thermal energy during heating, growth and cooldown.\\
Protocol 2, as displayed in Fig. \ref{Fig1}d), differed from Protocol 1 mainly by $P_{\mathrm{heat}}$ = $P_{\mathrm{cool}}$ = 1 $\times$ 10\ur{-4} mbar O\dr{2}. For this protocol, different $t_{\mathrm{pre}}^{\mathrm{ann}}$ durations were tested and initial flushing with O\dr{2} was excluded.\\
In Protocol 3, shown in Fig. \ref{Fig1}e), $T_{\mathrm{dep}}$ = 550 $^\circ$C and V2 and V3 left completely open without any O\dr{2} in-flow during the whole process except for the initial O\dr{2} flush. This decreases $P_{\mathrm{heat}}$ and $P_{\mathrm{cool}}$ to base pressure, which depend on the heating and pumping. \\
Protocol 4 was identical to Protocol 3 with the exception that $P_{\mathrm{cool}}$ = 10\ur{-4} mbar, as displayed in Fig. \ref{Fig1}f). The pressure peak initializing the cooldown corresponds to the PID fluctuating to find the correct value. It should be noted that, for all depositions, the pressure increases during ablation, since the ablated species from the target affect the pressure.\\
The \GAO growth thicknesses were measured by X-ray reflectivity (XRR) using a film grown by Protocol 1 ($d$\dr{\GAO} $\approx$ 8.5 nm). The growth time (4 min) was the same for all samples. X-ray diffraction (XRD), XRR, and atomic force microscopy (AFM), were used to assess the crystal structure.\\
Electrical connections were made by bonding with Al ultrasonic wire-bonding, in van der Pauw geometry. Electronic transport below room temperature was measured using a Cryogen-Free Measurement System equipped with a 16 T magnet.

\section{Indicative room temperature parameters}

\begin{figure}
    \centering
    \vspace{0pt}
    \includegraphics[width=.5\textwidth]{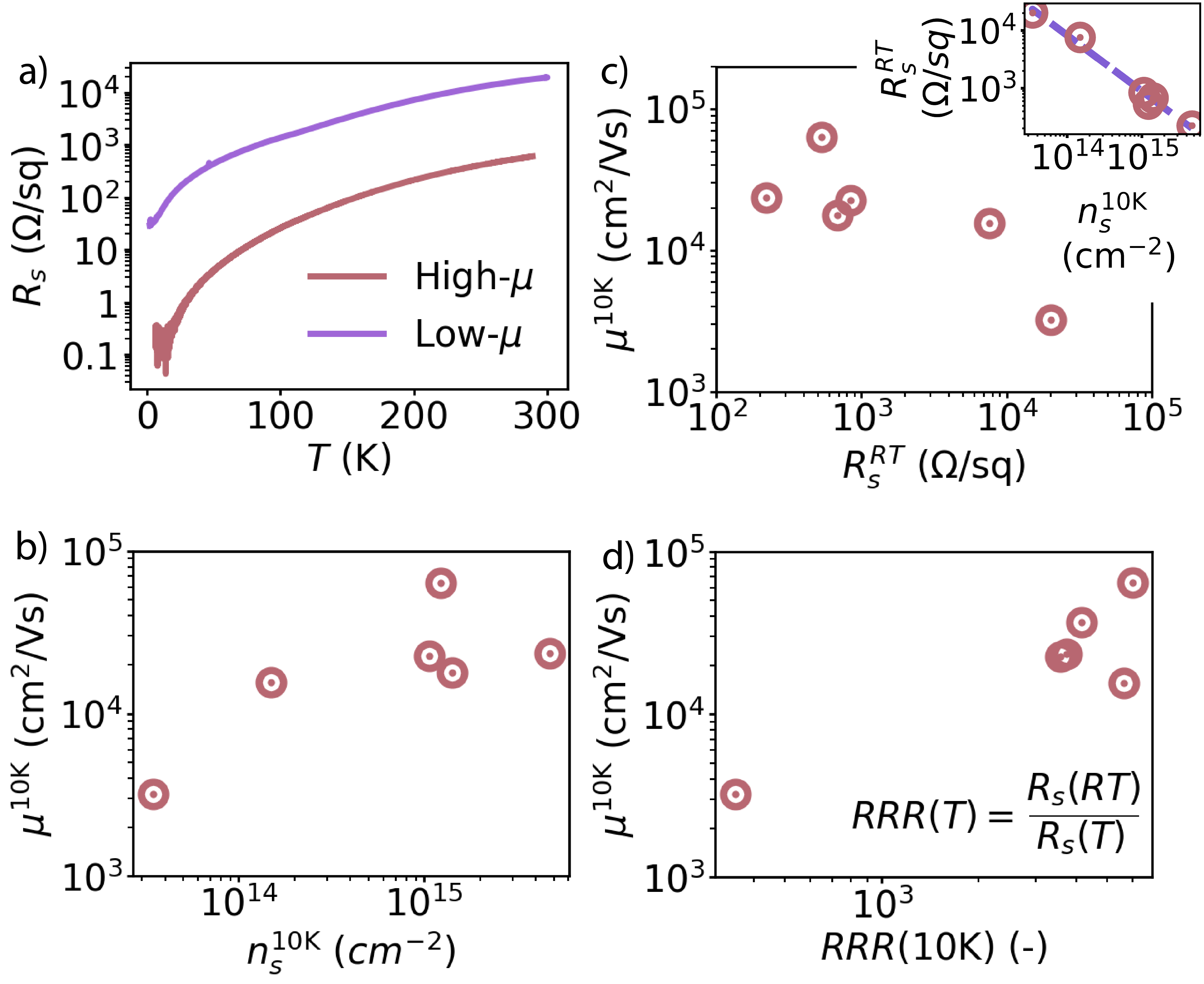}
    \caption{\textbf{Indicative values for high mobility prediction.} a) Sheet resistance, \textit{R}\d{\textit{s}}, as a function of temperature (\textit{T}) for a high mobility (brown) and a low mobility (purple) \GAO/\STO sample. The electron mobility at 10 K (\mgmu\u{\textit{10$\,$K}}) as a function of carrier concentration at 10$\,$K ($n_s^{10 \,K}$) in b), as a function of sheet resistance at room temperature ($R_s^{RT}$) in c), and as a function of residual resistance ratio (\textit{RRR}(10$\,$K)) in d). Inset in c) displays the $n_s^{10\,K}$-dependency of $R_s^{RT}$ The brown points are data and the purple dashed line a log-log linear fit. Carrier densities and mobilities are extracted through a single electronic band model as displayed in Supplementary Material Fig. S1.}
    \label{Fig3}
\end{figure}
\noindent
For the samples of this study, explicit low-temperature mobility measurements need cooling and testing in vdP and Hall configurations with magnetic field sweeps. Indicative room temperature parameters allow quicker mobility optimisation, as reported previously\cite{Christensen2018}. As displayed in Fig. \ref{Fig3}a) such indicative features are extractable from the sheet resistance temperature-dependency. Both high (4.2$\,\times \,10^4$ \mob at 2 K) and low (6$\,\times \,10^3$ \mob at 2 K) mobility samples here show metallic behaviour but differ in the absolute resistance and residual resistance ratio (\textit{RRR}(10 K) = \textit{R}\d{s}(300 K)/\textit{R}\d{s}(10 K)). 
The mobility as a function of carrier density is
shown in Fig. \ref{Fig3}b). A peak in mobility was reported\cite{Christensen2018} at $n_s \approx$ 4 $\times$ 10\ur{14} cm\ur{-2}, within a range of carrier densities not accessed in this investigation. Plotting the mobility as a function of room temperature sheet resistance (\RRT), displayed in Fig. \ref{Fig3}c), reveals a qualitative mirror image of Fig. \ref{Fig3}b). The inset indicates the log-log linear dependency between the carrier density and \RRT, explaining these similarities. The dependency between mobility and \RRT, through the carrier density, reveals \RRT as a relevant indicative value to estimate the low-temperature mobility.

\begin{figure*}
    \centering
    \vspace{0pt}
    \includegraphics[width=.8\textwidth]{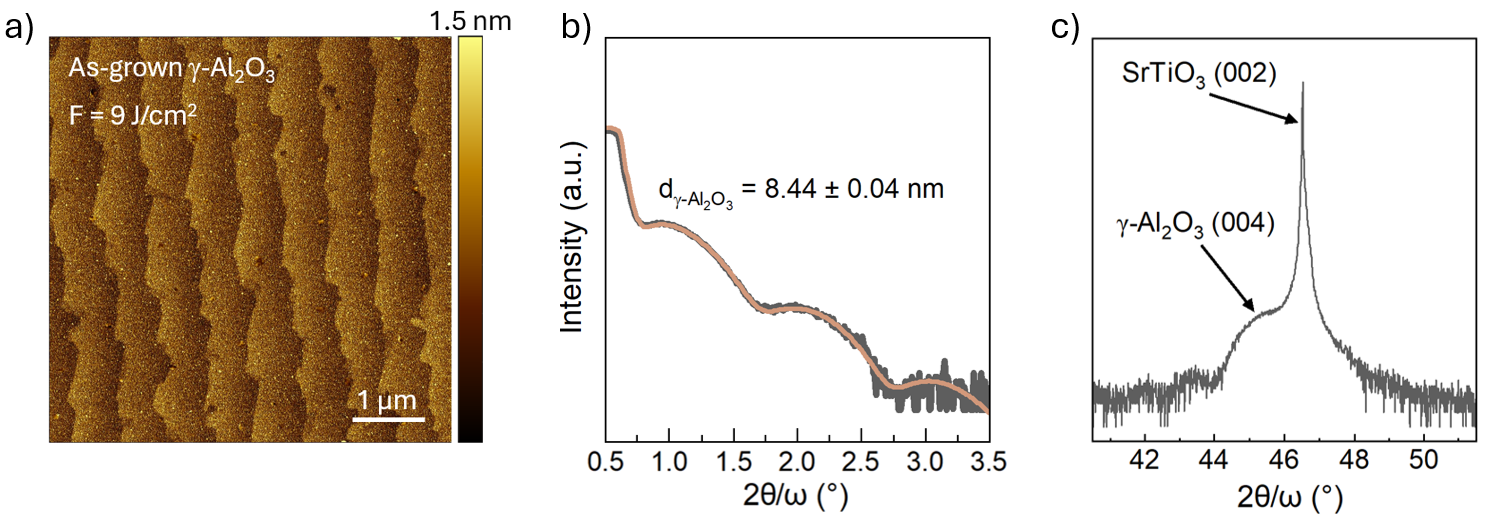}
    \caption{\textbf{Structural data of \GAO/\STO Sample 9.} a) AFM after \GAO deposition on \STO. b) X-ray reflectivity data and fit (solid brown line) yielding a \GAO thickness of $d_{\gamma\text{-}Al_2O_3} = 8.44 \,\pm\,0.4 $ nm. c) X-ray diffraction showing the \STO(002) and \GAO(004) peaks. Sample 9 was grown using protocol 1. The y-axes in b) and c) is logarithmic intensity.}
    \label{Fig2}
\end{figure*}
One can define a relevant upper threshold for high mobility in \RRT at 10 k$\Omega$/sq. A lower threshold is not determinable from Fig. \ref{Fig3}, although some level of conductance between the top and bottom of a sample can provide an alternative, as 3D conductance is undesirable for 2D features. A threshold 3D conductance requires additional investigations.\\
The total carrier density is often temperature independent, hence \textit{RRR} can be understood as an indirect measurement of the thermal contribution to $\mu$, which will be high for high $\mu(\mathrm{10K})$. Figure \ref{Fig3}d) exhibits the mobility at 10 K as a function of \textit{RRR}(10 K). \textit{RRR}(10 K) increases with increasing mobility, yielding a range for high-mobility (\gmu $>$ 10\ur{4}) samples at \textit{RRR}(10 K) $>$ 3 $\times$ 10\ur{3}. This can be observed because both \RRT and $\mu^\mathrm{10K}$ are proportional to the approximately temperature-independent $n_s$.

\section{Structural analysis}
To gain insights into the relations between the crystalline quality and the mobility, sample 9, grown with protocol 1, ($\mu^{10K} = 1.6 \times 10^4$ \mob) was measured using AFM, XRR, and XRD. The visible terraces in AFM after deposition, as displayed in Fig. \ref{Fig2}a), indicate crystalline and epitaxial growth of the \GAO thin film. It should be noted that the surface termination of the substrates before deposition did not show a clear single-termination. However, the high-mobility carriers in the $d_{xz/yz}$-band should reside deeper in the SrTiO$_3$ substrate than those in the $d_{xy}$-band,\cite{Chikina2021} why interface-related scattering from e.g. surface roughness should contribute less in $d_{xz/yz}$-dominated \GAO/SrTiO$_3$ compared to similar $d_{xy}$-dominated heterostructures. \\
Since reflection high-energy electron diffraction was not installed in this chamber, the thin film thicknesses and epitaxial crystalline growth were confirmed by XRD and XRR measurements. In Fig. \ref{Fig2}b), the XRR data is displayed along with a fit determining the \GAO thickness to $d_{\gamma\text{-}Al_2O_3} = 8.44 \pm 0.04 $ nm. Fig. \ref{Fig2}c) shows XRD 2$\theta/\omega$ scan of the \GAO sample 9 grown on \STO(001). \GAO grows with a positive out-of-plane strain of 0.721(1)\%. Calculated by using the \textit{c}-lattice parameter of bulk \GAO (7.911 Å).\cite{ZhouRongsheng1991}\\
\section{2DEG formation mechanisms}
\begin{figure*}[t]
    \centering
    \vspace{0pt}
    \includegraphics[width=.8\textwidth]{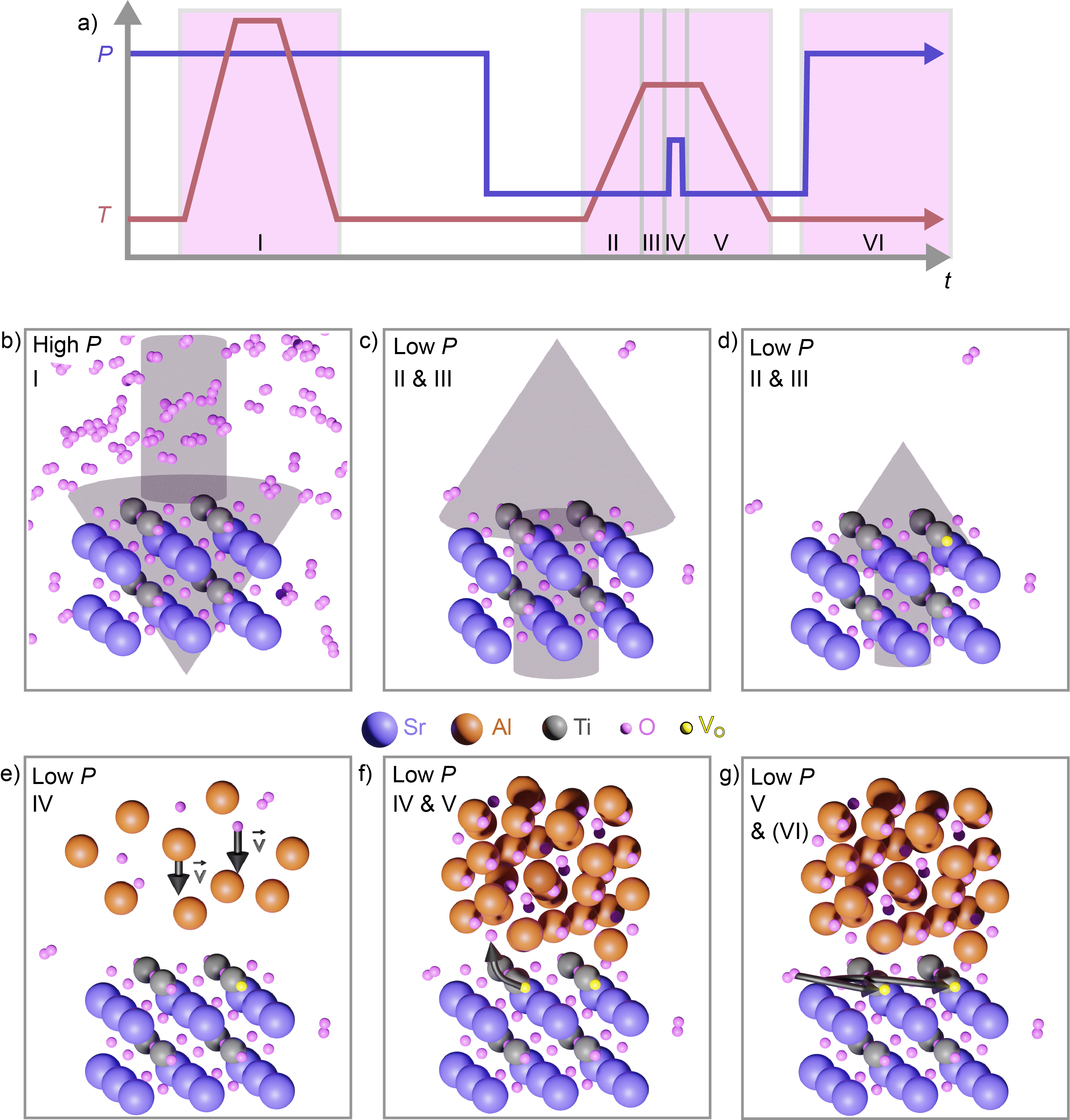}
    \caption{\textbf{Oxygen dynamics through the growth protocols.} a) Illustration of the full timespan of the growth process separated into six different regimes. b) \STO surrounded by oxygen molecules, as is the case during initial annealing. c) \STO in a low-pressure oxygen environment as the pulsed laser deposition chamber. d) same as c) but with an oxygen vacancy in the topmost layer. The arrows in b-d) indicates the direction and magnitude of the oxygen atoms diffusion to restore equilibrium, not drawn to scale. e) Bombardment of \STO in the initial part of the pulsed laser deposition. Here $\protect\vv{\mathrm{v}}$ is the velocity vector of the plasma species. f) Oxygen scavenging by the Al atoms of \GAO that have higher oxygen affinity than Ti and Sr in \STO. g) Oxygen recombination from the atmosphere. The arrows in f) and g) do not indicate forces but movements.}
    \label{Fig6}
\end{figure*}
\noindent
A thorough understanding of the 2DEG formation mechanism is essential to optimise the mobility in \GAO/\STO further. As previously described in ref. \cite{SC_Paper}, annealing  in oxygen causes a transition from \textit{R}\d{s}(RT) = 310 $\Omega$/sq to \textit{R}\d{s}(RT) $>$ 950 M$\Omega$/sq. This points towards oxygen vacancies being the primary source of the 2DEG formation.\\
Figure \ref{Fig6}a) illustrates how the samples go through six regimes in the growth process. Regime I involves annealing at atmospheric pressure and high temperatures. The sample is then loaded into the PLD chamber, initiating Regime II during heating under low pressure. Regime III occurs as the \STO surface remains exposed to low pressure at the growth temperature. Deposition begins in Regime IV and ends before Regime V, which includes cooldown. Regime VI covers subsequent sample storage.\\
Regime I is illustrated in Fig. \ref{Fig6}b), where the arrow indicates the direction and magnitude of the oxygen atom diffusion to restore thermodynamic equilibrium between the \STO lattice and the environment. 
Since the temperature reaches 1000\ur{$\;$o}C in O\dr{2} environment, the oxygen atoms become highly mobile and vacancies are filled. This creates a reproducible starting point regarding oxygen distribution. During the subsequent cooling, the atomic mobility decreases, making the oxygen occupation in the sample reflect the environment, functioning as a pressure memory of the furnace equilibrium.\\
In regimes II and III, illustrated in Fig. \ref{Fig6}c), the substrate is loaded in the PLD chamber, where the pressure is adjusted to $\leq$ 10\ur{-4} mbar. Now, the pressure balance is reversed. With the low-pressure environment and the memory of the furnace pressure in the lattice, the oxygen atoms will diffuse out of the lattice to restore thermodynamic equilibrium. The topmost atomic layer will experience the fastest reduction since this layer is directly exposed to the environment.
Once a vacancy is generated in the topmost layer, the system has moved slightly towards equilibrium. This decreases the oxygen diffusion, as illustrated in Fig. \ref{Fig6}d) and allows the oxygen vacancies to move further into the bulk \STO,. 
When the substrate reaches deposition temperature, a laser pulse is sent into the target, entering regime IV. The laser pulse is ablated into a high-energy plasma. When the plasma reaches the substrate a bombardment effect occurs, as illustrated in Fig. \ref{Fig6}e). The plasma adds energy on the scale of tens of eV\cite{Sambri2012,von_soosten2019} to the top layers of the lattice, enabling the escape of light elements. In \STO, the light element is oxygen.\\
Because \GAO only possess a single-element cation, one can increase the fluence without considering the effect on the cationic stoichiometry. Hence, one can utilise the fluence to access a larger range of energetic minima associated with crystal growth of \GAO compared to e.g. \LAO or LaTiO$_3$.\\
As the \GAO thin film is grown in the oxygen-deficient environment, it further reduces \STO since the oxygen affinity of Al is higher than that of Ti and Sr. The combination of the high temperature and the kinetic energy from the plasma ensures high atomic mobility.\cite{Chen2011,Christensen2017_2,von_soosten2019} This corresponds to regime IV and V and is illustrated in Fig. \ref{Fig6}f).\\
At last, during sample cooldown (regime V), for Protocol 2 and 4, the pressure is increased from deposition pressure to 10\ur{-4} mbar. This may cause oxygen recombination, as displayed in Fig. \ref{Fig6}g), where an atmospheric oxygen molecule is split up and added to the lattice, annihilating two vacancies.\cite{Christensen2017_2,von_soosten2019,Hvid-Olsen2022} This effect may also occur during subsequent storage corresponding to regime VI.\\ 
\section{Protocol comparison}
Samples 1, 2, 3, 9, and 10 grown by Protocol 1, as already displayed in Fig. \ref{Fig3}, were confirmed to host high-mobility 2DEGs. Protocol 1, however, has some weaknesses. Stabilising a constant low pressure during the temperature profile is challenging, often failing to reach the intended low pressure during heating, probably due to outgassing from the silver paint and heater. This is likely the reason for the large fluctuation in \RRT, as shown in Supplementary Material Table S1. Additionally, manual pressure control is time-consuming ($>$5 hours) and less reproducible than automated control. \\
To increase the reproducibility and time efficiency of the growth, Protocol 2, with $P_\mathrm{heat}=P_\mathrm{cool}=1 \times 10^{-4}$ mbar, was attempted to create samples 19-33. In contrast to Protocol 1, however, Protocol 2 shows measurable conductance \textit{R}\ur{2G}(RT) at room temperature between the chip carrier back plane and any of the four corners on the upper surface of the sample. Thus, Protocol 2 yields bulk conductance before high mobility 2DEGs, when decreasing the sheet resistance by extending $t_{\mathrm{pre}}^\mathrm{ann}$.\\
The tendency to create bulk conductance before high-mobility 2DEGs, could be explained by the balance between the effects of temperature and pressure. When the substrate is uncapped by the thin film, in regimes II and III, the pressure primarily affects the surface of the sample, where the pressure gradient is highest. In contrast, the temperature acts primarily on the back side of the substrate, where it is thermally anchored and then decreasingly through the bulk and up to the front surface. The pressure gradient at the surface effectively pulls the oxygen atoms out of the lattice, while the temperature facilitates energy to allow atomic movement. Thus, both effects are needed, but the low pressure drives the surface oxygen vacancy formation, while the temperature drives a homogenous distribution of oxygen vacancies. Therefore, a dominating low-pressure before deposition is preferential for high-mobility 2DEGs.\\
Attempting to leverage this mechanism for the formation of surface oxygen vacancies, Protocol 3 was applied for growing samples 34-36, as shown in Table S1. This protocol also yields 3D conductance before high mobility 2DEG formation. In this case, however, the 3D conductance likely develops during post-annealing and cooldown when the pressure is at its lowest, the temperature is still significant, and the potential 2DEG area is shielded by the \GAO, acting as a capping layer that keeps the oxygen from escaping from the 2DEG area.\\
To counteract both the lack of surface oxygen vacancies from Protocol 2 and the increase in bulk oxygen vacancies in Protocol 3, Protocol 4 was developed as displayed by samples 37-40. Here, the bulk conductance does not reach measurable ranges, but the 2DEG resistance tends to increase with increasing growth temperature, rendering the protocol unsuited for high-mobility 2DEG growth. This can be attributed to oxygen recombination while the pressure is 10\ur{-4} mbar  and during the peak in pressure while the pressure PID fluctuates towards the correct value.\\
\section{Conclusion}
Within this investigation, four different growth protocols attempting to grow \GAO on \STO are presented alongside small variations within each of them. The purpose was to create a 2DEG with high carrier mobility. This purpose was initially reached by Protocol 1, displaying a growth window of such epitaxial, crystalline, high mobility \GAO/\STO 2DEGs. Protocol 1 however, is highly dependent on manual work, decreasing the reproducibility. Hence, the subsequent protocols attempted to stay within the high mobility growth window while automating the work. To optimise that process, the residual resistance ratio (\textit{RRR}) and the room-temperature sheet resistance (\textit{R}\d{s}) were identified as indicative parameters for estimating the mobility at 10 K without time-consuming low-temperature measurements between every growth. For both \textit{RRR} and \textit{R}\d{s}, the reason for the indicative implications is identified as the approximate temperature independence of the carrier concentration of the \GAO/\STO 2DEGs.
Four different mechanisms for oxygen-vacancy-driven 2DEG formation were discussed. These were \STO-environment equilibrium, bombardment, reduction of \STO by \GAO, and oxygen recombination. \GAO having a single-element cation allows high laser fluences during growth without affecting stoichiometry, providing an increased bombardment effect and available energy for the crystal growth. Considering the growth mechanisms and comparing 40 samples grown with the four different protocols, pointed toward a
need for low pressure to induce oxygen vacancies in the surface layers of \STO before deposition.\\
\section{Supplementary material}
The supplementary material includes an overview of the 40 samples described in this investigation, linear fitting of the Hall resistance used to extract the carrier densities and mobilities in Fig. \ref{Fig2}. Furthermore, it includes a detailed electronic transport analysis of one high-mobility sample (sample 1), including a 2-band and anomalous hall effect fitting.
\section{Acknowledgements}
We thank R. T. Dahm, C. N.  Lobato, V. Rosendal, and C. E. N. Petersen for helpful discussions.
T.H.O., A. C. and F.T. acknowledge support by research grant 37338 (SANSIT) from Villum Fonden. T.S.J. acknowledges support from the Novo Nordic Foundation Challenge Program, grant no. NNF21OC0066526 (BioMag).


%
%

%



\bibliography{aipsamp}

\end{document}